\documentclass[twocolumn,prl,aps,longbibliography,superscriptaddress,preprintnumbers,tightenlines,showpacs,nofootinbib,amsfonts,amsmath]{revtex4-2}

\pdfoutput=1

\usepackage{epsfig}
\usepackage{graphics}
\usepackage{graphicx}
\usepackage{amsmath,amssymb,mathrsfs}
\usepackage{amsfonts}
\usepackage[usenames,dvipsnames]{xcolor}
\usepackage{xcolor}
\usepackage{wasysym}
\usepackage{times}
\usepackage{newtxmath}
\usepackage{gensymb}
\usepackage{appendix}
\usepackage{listings}
\usepackage{url}
\usepackage[normalem]{ulem}
\usepackage{alltt}
\usepackage[colorlinks=true,
urlcolor=blue]{hyperref}
\usepackage{cleveref}
\usepackage{longtable}
\usepackage{enumitem}
\setlist{nosep}
\usepackage{color}
\usepackage{calc}
\usepackage{tensor}
\usepackage[export]{adjustbox}
\usepackage{bm}
\usepackage{times}
\usepackage{multirow}
\usepackage{float}
\usepackage{dcolumn}
\usepackage[nolist,nohyperlinks]{acronym}
\usepackage{xspace}
\usepackage[english]{babel}
\usepackage[abs]{overpic}
\usepackage{pict2e}
\allowdisplaybreaks[1]
\usepackage[utf8]{inputenc}
\usepackage{gensymb}
\usepackage{bm}
\usepackage{stackengine}
\usepackage{boldline,multirow}
\usepackage{braket}
\usepackage{longtable}
\usepackage{orcidlink}

\usepackage{tabularx}
\usepackage{rotating}

\Crefname{figure}{Fig.}{Figs.}

\def\mR{\mathcal{R}}

\definecolor{dodgerblue}{HTML}{1E90FF}
\definecolor{viennared}{HTML}{DA0A14}

\hypersetup{citecolor=dodgerblue}

\graphicspath{{Figures/}}

\begin{document}

\title{
Relieving scale disparity in binary black hole simulations 
}

\newcommand{\aei}{\affiliation{Max Planck Institute for Gravitational Physics
	(Albert Einstein Institute), Am M{\"u}hlenberg 1, 14476 Potsdam, Germany}}
\newcommand{\southampton}{\affiliation{School of Mathematical Sciences and STAG Research Centre, University of Southampton, Southampton, SO17 1BJ, United Kingdom}}
\newcommand{\CornellPhysics}{\affiliation{Department of Physics, Cornell University, Ithaca, NY, 14853, USA}}
\newcommand{\Cornell}{\affiliation{Cornell Center for Astrophysics and Planetary Science, Cornell University, Ithaca, New York 14853, USA}}
\newcommand{\CornellLep}{\affiliation{Laboratory for Elementary Particle Physics, Cornell University, Ithaca, New York 14853, USA}}
\newcommand{\Caltech}{\affiliation{Theoretical Astrophysics 350-17, California Institute of Technology, Pasadena, CA 91125, USA}}
\newcommand{\Fullerton}{\affiliation{Nicholas and Lee Begovich Center for Gravitational-Wave Physics and Astronomy, California State University Fullerton, Fullerton, CA 92834, USA}}

\author{Nikolas A. Wittek \orcidlink{0000-0001-8575-5450}} \aei
\author{Leor Barack \orcidlink{0000-0003-4742-9413}} \southampton
\author{Harald P. Pfeiffer \orcidlink{0000-0001-9288-519X}} \aei
\author{Adam Pound \orcidlink{0000-0001-9446-0638}} \southampton
\author{Nils Deppe \orcidlink{0000-0003-4557-4115}} \CornellLep \CornellPhysics \Cornell
\author{Lawrence E.~Kidder \orcidlink{0000-0001-5392-7342}} \Cornell
\author{Alexandra Macedo \orcidlink{0009-0001-7671-6377}} \Fullerton
\author{Kyle C.~Nelli \orcidlink{0000-0003-2426-8768}} \Caltech
\author{William Throwe \orcidlink{0000-0001-5059-4378}} \Cornell
\author{Nils L.~Vu \orcidlink{0000-0002-5767-3949}} \Caltech

\date{\today}

\begin{abstract}
{\it Worldtube excision} is a method of reducing computational burden in Numerical Relativity simulations of binary black holes in situations where there is a good analytical model of the geometry around (one or both of) the objects. Two such scenarios of relevance in gravitational-wave astronomy are (1) the case of mass-disparate systems, and (2) the early inspiral when the separation is still large. Here we illustrate the utility and flexibility of this technique with simulations of the fully self-consistent radiative evolution in the model problem of a scalar charge orbiting a Schwarzschild black hole under the effect of scalar-field radiation reaction. We explore a range of orbital configurations, including inspirals with large eccentricity (which we follow through to the final plunge and ringdown) and hyperbolic scattering. 
\end{abstract}

\maketitle
\acrodef{NR}{numerical relativity}
\acrodef{GW}{gravitational wave}
\acrodef{BBH}{binary black hole}
\acrodef{BH}{black hole}

\newcommand{\NR}[0]{\ac{NR}\xspace}
\newcommand{\BBH}[0]{\ac{BBH}\xspace}
\newcommand{\BH}[0]{\ac{BH}\xspace}

\newcommand{\citeme}[0]{{\color{purple}{Citation!}}}

{\it Introduction}---%
The LIGO-Virgo-KAGRA (LVK) public catalog of
gravitational-wave transients lists around a hundred candidate events
from merging compact binaries consisting primarily of black holes
(BHs)~\cite{KAGRA:2021vkt}. Over 200 events are expected to have been observed
by the end of the fourth observing run in 2025~\cite{KAGRA:2013rdx}. These
observations begin to reveal
a population of compact binaries with a
large mass asymmetry. The most extreme example so far is 
candidate event ${\rm GW}191219{\_}163120$, with inferred component
masses of $31.1^{+2.2}_{-2.8}M_\odot$ and $1.17^{+0.07}_{-0.06}M_\odot$, suggestive of BH and neutron star
progenitors~\cite{KAGRA:2021vkt}. The LVK analysis notes that the
source's mass ratio of $\sim\! 1\,{:}\,26$ ``is extremely challenging
for waveform modeling,\; \ldots{} with the bulk of the posterior
probability distribution [lying] outside the range of calibration of
the waveforms'' and cautions about systematic
uncertainties in results for this candidate.  Indeed, we are reaching
a situation where the quality of data extractable from
gravitational-wave measurements is in some scenarios dictated by the
limited accuracy of waveform models~\cite{Dhani:2024jja}, diminishing the
return from technological advance. The problem will inevitably become
increasingly more acute as detector sensitivity improves and as we
probe deeper into poorly modeled regions of the binary parameter
space~\cite{Purrer:2019jcp,Hu:2022rjq,Hu:2022bji,Jan:2023raq,Owen:2023mid,Kapil:2024zdn}. The problem of modelling mass-asymmetric mergers stands out as
particularly
urgent \cite{LISAConsortiumWaveformWorkingGroup:2023arg,Pompili:2023tna}.
\begin{figure}
\parbox[t]{0.45\linewidth}{
  \includegraphics[width=0.84\linewidth, trim = 4.4cm 24.2cm 11cm 0cm, clip, valign=t]{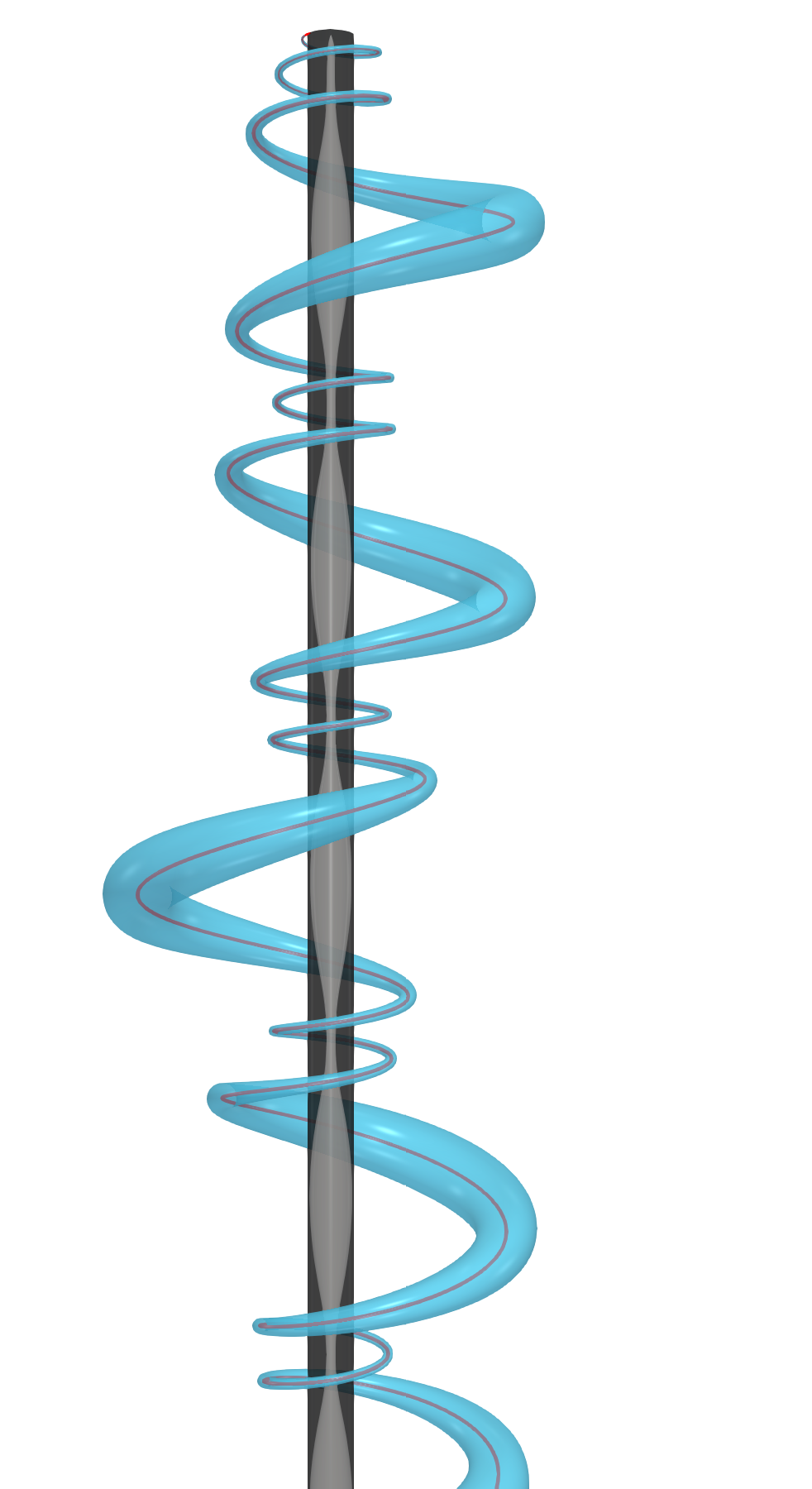}
}
\hspace*{-.2cm}
\parbox[t]{0.55\linewidth}{
  \vspace*{.0cm}
  \includegraphics[width=\linewidth, trim = 12cm 0 75cm 5cm, clip, valign=t]{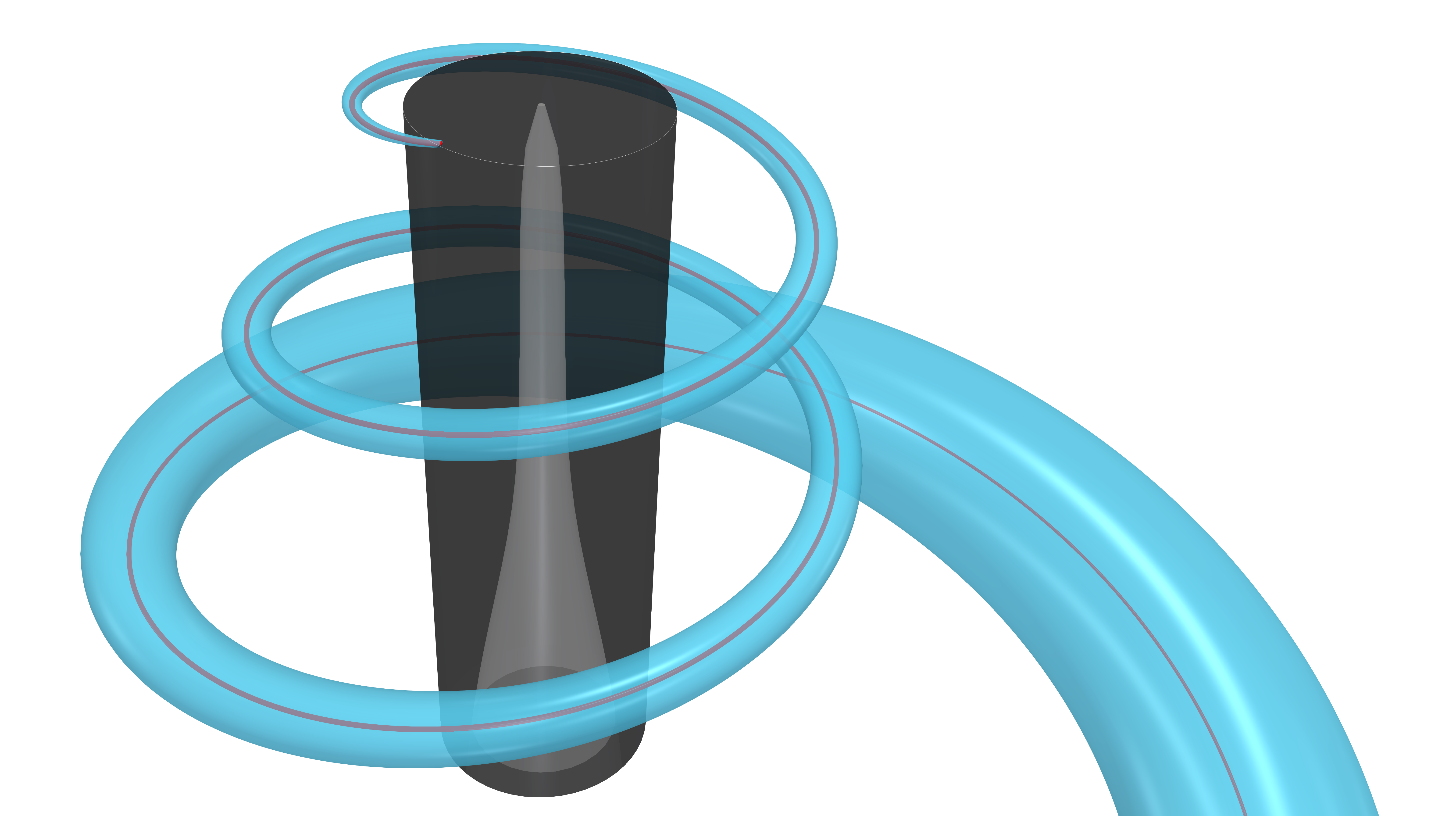}\\
  \vspace*{.0cm}
\hspace*{-1cm}\includegraphics[width=0.56\linewidth, trim=1.7cm 28.5cm 54cm 0, clip]{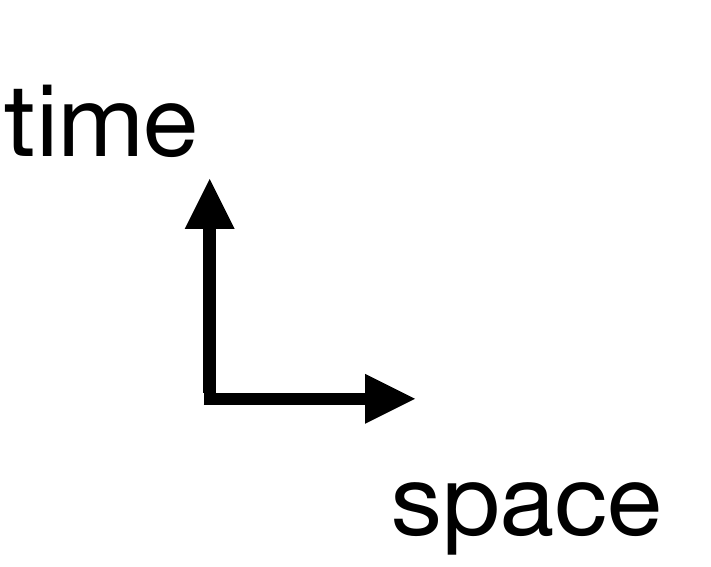}
 }
  \caption{\label{fig:fancy_pic} Illustration of worldtube
      excision.  The left shows a space-time diagram of an eccentric
      inspiral.  The red line represents the trajectory of the small
      object orbiting a black hole represented by the black cylinder.
      The cyan region around the trajectory represents the worldtube,
      within which a perturbative analytical solution is employed.
      The field equations outside the worldtube are solved with full 3+1 dimensional      NR, extending through the BH horizon to an
      inner excision boundary shown in grey.  The right shows an
      enlargement of the last few orbits and merger.}
\end{figure}

Numerical Relativity (NR) simulations of mass-asymmetric binaries are challenging due to the {\it scale disparity} inherent in the problem: One has to resolve small-scale features associated with the lighter object concurrently with features whose lengthscale is set by the larger (or total) mass---like gravitational waves near merger.  This leads to a stringent demand on the time resolution of the simulation, dictated by the Courant-Friedrichs-Lewy (CFL) condition. To make matters worse, the number of observable radiative inspiral cycles increases
with mass asymmetry, necessitating a longer evolution time. In terms of the mass ratio $q\leq 1$, the
combined effect is a computing cost that scales approximately like $q^{-2}$~\cite{Dhesi:2021yje}, in practice making full simulations prohibitive for $q$ much smaller than $\sim1/10$. The most ambitious simulation so far tracked the last 13 orbital cycles prior to merger of a system with $q=1/128$~\cite{Lousto:2020tnb, Rosato:2021jsq}, with even smaller mass ratios attempted in head-on collision scenarios~\cite{Sperhake:2011ik,Lousto:2022}. However, the computational cost of such simulations remains extremely high, and they are yet to be developed for long inspirals or to include essential astrophysical features such as spin and orbital eccentricity.

Gravitational self-force theory~\cite{Barack:2018yvs} does offer an alternative tailored to the small-$q$ regime, and it can potentially cover a large part of the parameter space that is beyond the reach of current NR methods~\cite{Albertini:2022rfe}. However, state-of-the-art ``post-adiabatic'' self-force waveforms that achieve high accuracy are currently only available in the special case of quasicircular orbits of a small compact object around a slowly spinning large black hole~\cite{Wardell:2021fyy,Mathews:2024}. Extensions to eccentric, precessing orbits, rapid spin, and through the final merger-ringdown phase will likely require a decade of development. The accuracy  such waveforms will achieve in these less specialized scenarios is also unknown.

Our {\it worldtube excision} approach aims to provide a systematic mitigation of the small-$q$ problem, by directly alleviating scale disparity in NR binary simulations. The basic idea is simple: A large region around the smaller object is excised from the numerical domain, and the spacetime metric inside it is replaced with an approximate analytical expression (e.g., one representing a tidally perturbed BH geometry). The smallest lengthscale is now that of the excised sphere (a “worldtube” in spacetime), instead of the scale of the smaller body. The 
CFL stability limit on the timestep of the numerical simulation is relaxed, allowing a commensurate reduction in run time. 
In principle, worldtube excision can be usefully applied in any situation where a good analytical approximation for the metric around the smaller (or both) objects is available, including the early inspiral stage where the gravitational interaction is relatively weak. In this letter, however, we maintain focus on the case of near-merger binaries with a large mass disparity, which provides our main motivation.

The basic construction is illustrated in Fig.~\ref{fig:fancy_pic},
displaying actual results from one of our NR simulations, as detailed below.
  The spacetime diagram shows the
eccentric inspiral orbit of the small object (red trajectory) around
the worldtube of the larger object (a BH, in black), with the blue
region representing the excised worldtube, whose radius is dynamically
adjusted in accordance with a certain algorithm to be explained
below. This simulation tracks the inspiral all through to the final
merger, zoomed onto in the right panel.

Initial development of the worldtube excision technique began in Ref.~\cite{Dhesi:2021yje} and was later implemented and tested using simple configurations in Refs.~\cite{Wittek:2023nyi,Wittek:2024gxn}. Here we report a culmination of this program, in the form of a versatile NR implementation on the SpECTRE platform~\cite{spectrecode}, whose utility and flexibility we illustrate with fully self-consistent evolution simulations for a range of orbital configurations. These include inspirals with eccentricity as high as $e\!\sim\! 0.9$, which we follow through to merger and ringdown, as well as hyperbolic scattering with large scattering angles---a scenario of much recent interest, e.g.~\cite{Damour:2016gwp,Bern:2019nnu,Kalin:2019rwq,Kalin:2020fhe, Khalil:2022ylj,Rettegno:2023ghr,Adamo:2024oxy,Fontbute:2024amb,Jakobsen:2023ndj,Barack:2023oqp,Gonzo:2024xjk,Driesse:2024xad,Long:2024ltn}.

As in \cite{Dhesi:2021yje,Wittek:2023nyi,Wittek:2024gxn}, our binary system is made up of a Schwarzschild BH (the ``large object'') and a pointlike, nonspinning scalar charge of negligible gravitational mass. The particle is assumed to source a massless, minimally-coupled scalar field, and the back-reaction from the scalar field onto the charge drives the radiative inspiral; gravitational back-reaction is ignored for simplicity. This setting retains many of the pertinent challenges of the astrophysical compact binary problem, while postponing the need to tackle the fully nonlinear Einstein's field equations. At the end of this Letter we preview initial work towards the final step of replacing the scalar charge with a small BH, our ultimate goal.

Even within the scalar problem, our implementation provides a vital benchmark for traditional self-force calculations. We present the first (to our knowledge) fully
self-consistent solutions of the BH-scalar-charge
inspiral-merger problem with eccentricity (extending the special case of
quasi-circular inspirals in \cite{Wittek:2024gxn}). They
are fully {\it self-consistent} in that, unlike the standard waveform-generation frameworks used in
self-force theory \cite{Chua:2020stf,Miller:2020bft,Hughes:2021exa,Katz:2021yft,Isoyama:2021jjd,Pound:2021qin,McCart:2021upc,Drummond:2023wqc,Nasipak:2023kuf}, they do not rely on a
two-timescale expansion (which breaks down prior to merger), nor do they rely on an expansion around a reference geodesic (which cannot capture large qualitative changes in behavior). Our method does incur error from the finite accuracy of the analytical model applied inside the
worldtube, but this error is readily controlled by
extending the analytical approximation to higher order or reducing the worldtube radius~\cite{Wittek:2023nyi,Wittek:2024gxn}. We expect these dual advantages of flexibility and controllable error to persist in the fully nonlinear gravitational case.

\begin{figure*}[t]
	\includegraphics[width=6.8cm,clip=true,trim=15 25 0 10]{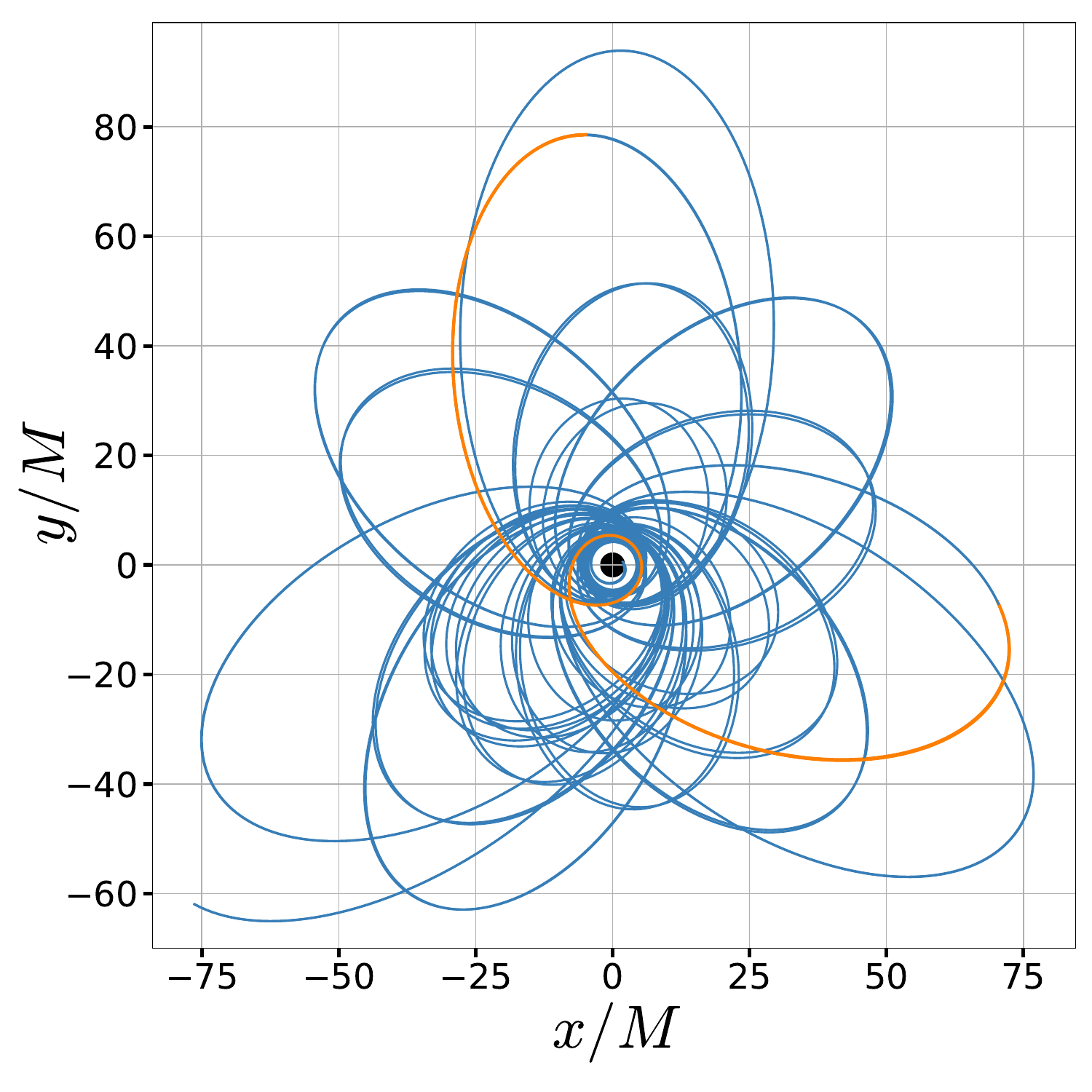}
	\includegraphics[width=10.7cm, trim = 0 18 20 20]{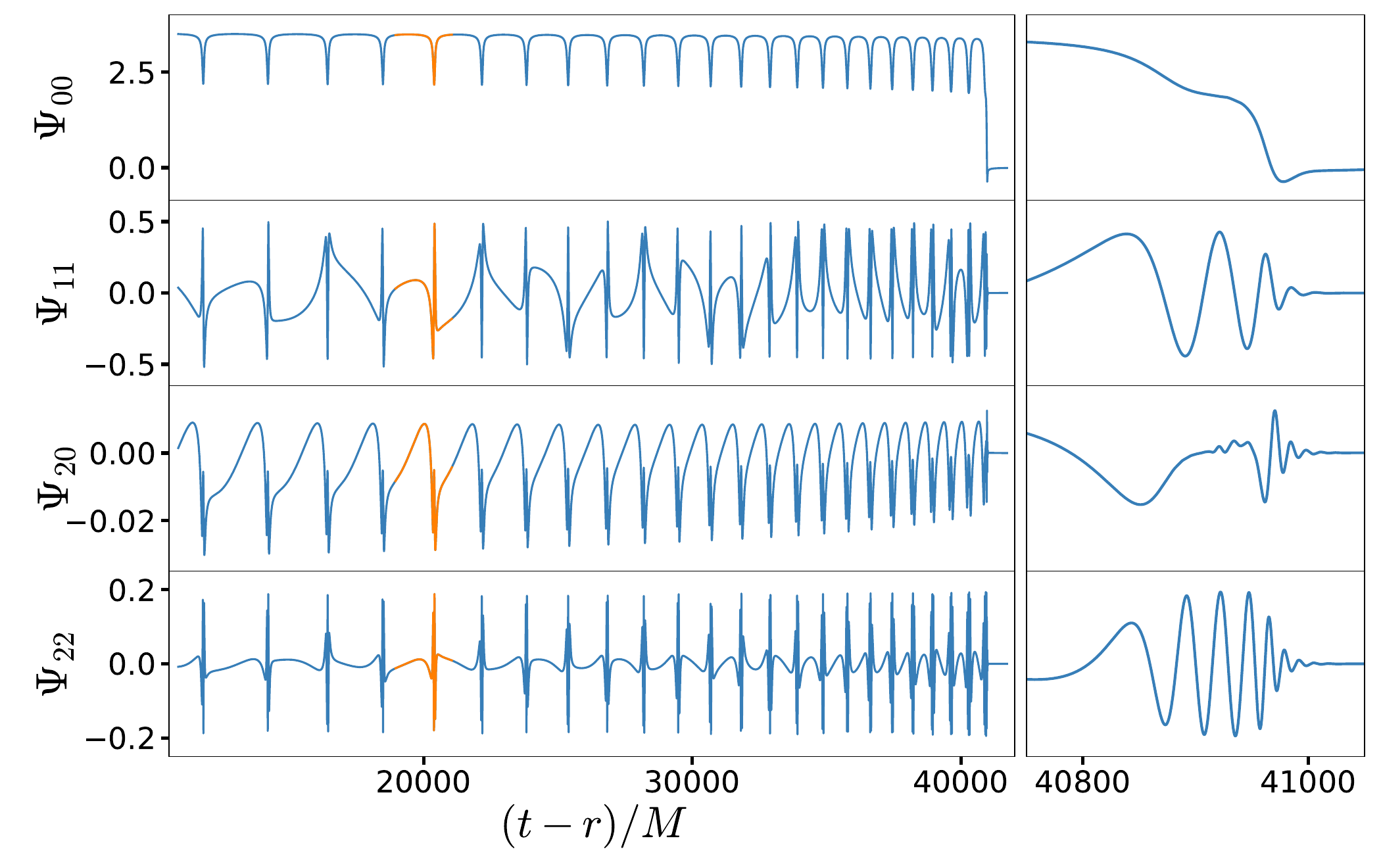}
	\caption{Simulation of an eccentric binary.  \textbf{Left:} orbital trajectory for the simulation starting from the point at which the self-force is turned on. \textbf{Right:} some important waveform modes. Highlighted in orange is one radial period in both panels. The ringdown is shown enlarged in the rightmost panel. This simulation starts with initial eccentricity $e=0.9$ at apoapsis, $r=100M$, and uses $\epsilon=0.02$.
	}
	\label{fig:ecc_xy}
\end{figure*}
{\it Binary model}---%
We consider a Schwarzschild BH of mass $M$, orbited by a pointlike particle carrying a charge $Q$ and mass $\mu\ll M$. The particle sources a test scalar field $\Psi$, assumed to satisfy the massless Klein-Gordon equation 
\begin{equation}\label{KG}
g^{\mu\nu}\nabla_{\!\mu}\nabla_{\!\nu}\Psi\;=\,-4\pi Q \int\! \delta^4(x^\alpha-x_{p}^\alpha(\tau))(-g)^{-1/2}d\tau,
\end{equation}
with boundary conditions corresponding to no radiation coming out of the BH or in from infinity. 
The background geometry is represented by the Schwarzschild metric $g_{\mu\nu}$, with inverse $g^{\mu\nu}$, determinant $g$, and compatible covariant derivative $\nabla_{\!\mu}$.
The particle's worldline $x^\alpha_p(\tau)$ is
parametrized by proper time $\tau$, and throughout this letter we use units in which $G=1=c$.
In this setup, the dimensionless small parameter is $\epsilon:= Q^2/(\mu M)\ll 1$, analogous to $q\ll 1$ in the binary BH problem.
In the limit $\epsilon\to 0$ (with $\mu/Q\to\text{constant}$), $x^\alpha_{p}(\tau)$ describes a geodesic orbit in the Schwarzschild geometry of the BH, whereas for finite $\epsilon$, the particle experiences a weak self-force due to back-reaction from $\Psi$, which slowly accelerates it away from geodesic motion.
Concretely, the particle's equation of motion reads~\cite{Quinn_2000,Poisson:2011nh,Harte:2014wya}
\begin{equation}\label{EOM}
u^\beta \nabla_{\!\beta} \left( \mu u^\alpha \right) = Q \nabla^\alpha \Psi^\mR,
\end{equation}
where $u^\alpha:=dx_p^\alpha/d\tau$, and $\Psi^\mR$ is the Detweiler-Whiting regular piece of $\Psi$~\cite{Detweiler:2002mi} (a certain smooth function satisfying $g^{\mu\nu}\nabla_{\!\mu}\nabla_{\!\nu}\Psi^\mR\!=\!0$), whose gradient here is evaluated at the particle. Note Eq.~\eqref{EOM} implies $d\mu/d\tau\! =\! Q(d\Psi^\mR/d\tau)$, with variation in the particle's rest mass representing exchange of energy with the scalar field.

To obtain the binary evolution and scalar-wave emission, one's task is to solve the coupled equations (\ref{KG}) and (\ref{EOM}), with the above boundary conditions for $\Psi$ and given initial values for $x^\alpha_{p}$ and $u^\alpha$. 

 {\it Worldtube excision method}---%
Introducing standard Kerr-Schild (KS) coordinates on the Schwarzschild background, we define the worldtube $\Gamma$ as a KS coordinate sphere centered on the charge on each hypersurface of constant KS time $t$. In Ref.~\cite{Wittek:2024gxn} we constructed an approximate analytical solution to (\ref{KG}), valid in the vicinity of the charge inside $\Gamma$, for a generic $x^\alpha_p(\tau)$. 
The approximation has the form $\Psi^{\cal P}+\Psi^\mR$, where the ``puncture field'' $\Psi^{\cal P}$ is an approximate particular solution of the inhomogeneous equation (\ref{KG}), given explicitly in \cite{Wittek:2024gxn} as a truncated expansion of the Detweiler-Whiting singular field~\cite{Detweiler:2002mi} in powers of coordinate distance from the charge; analogously, we approximate the smooth field $\Psi^\mR$ with a truncated Taylor expansion.
Our analytical approximant thus contains a set of a-priori unknown coefficients coming from the Taylor expansion of $\Psi^{\cal R}$ and its time derivative at the charge. An NR evolution in KS time $t$ is then set up (see below), which matches 
the numerical solution outside $\Gamma$ to the analytical field on $\Gamma$ mode by mode in a multipole expansion around the particle.  The matching fixes the unknown coefficients in the analytical field, and hence also the value of $\nabla_{\!\beta}\Psi^R$ at the charge. 
This, in turn, determines the self-forcing term in (\ref{EOM}), allowing us to evolve the worldline in time, as well as providing boundary data  to the NR evolution outside the worldtube. See \cite{Wittek:2024gxn} for full detail.

A complication is that $\Psi^{\cal P}$ itself depends on the charge's self-acceleration, which is a-priori unknown, leading to implicit matching conditions. We resolve this by applying an iteration procedure using the small magnitude of the self-acceleration as a perturbative parameter \cite{Wittek:2024gxn}. This introduces additional error, but one that is easily controllable and made subdominant in practice.

{\it NR method}---%
Our method is implemented in SpECTRE~\cite{spectrecode}, which employs a Discontinuous Galerkin (DG) method to evolve Eq.~\eqref{KG}. In standard black hole excision~\cite{Scheel:2006gg,Hemberger:2012jz,Scheel:2014ina,Lovelace:2024wra}, the motion of the excision spheres in the computational domain is controlled using a series of time-dependent maps which track the apparent horizons. We use this infrastructure to adjust the time-dependent maps at each time step according to the acceleration of the pointlike particle obtained from Eq.~\eqref{EOM}.

The choice of worldtube radius trades off between computational saving (larger $R$) and accuracy of the analytical approximation (smaller $R$). In the case of quasicircular orbits, the choice $R(t)\propto r_p^{3/2}(t)$ (where $r_p(t)$ is the KS orbital radius at time $t$) is motivated from an examination of how the estimated error in our local approximation to $\Psi$ on $\Gamma$ depends on $r_p$ \cite{Wittek:2024gxn}.
Once the worldtube radius becomes so large that the CFL limit is set by a different part of the computational domain (like the vicinity of the BH of size $\sim M$), there is no benefit from increasing it further.  Therefore,
for the eccentric and scattering orbits considered in this work,
we introduce an upper bound $R_{\infty}$ for the worldtube radius by taking
\begin{equation}\label{eq:wt_radius_func}
	R(t)= R_\infty \left(\frac{r_p(t)}{r_0}\right)^{3/2}\left[ \left( 1 + \left(r_p(t)/r_0\right)^{1 / \Delta}\right)\right]^{-3\Delta/2},
\end{equation}
which transits smoothly between $R(t)\sim r_p^{3/2}(t)$ for $r_p \ll r_0$ and $R(t)\sim {\rm const}=R_{\infty}$ for $r_p\gg r_0$.
The simulations presented here use $R_\infty\!=\!3M$ and a transition parameter $\Delta\!=\!0.05$; $r_0$ is chosen to achieve a worldtube radius $R_{6M}$ at the distance $r_p\!=\!6M$, typically $R_{6M}=0.2$--$0.8M$.
Adjustments were made also to the DG domain itself by tuning the fixed polynomial order of the elements to accommodate large changes in grid spacing for highly eccentric orbits and hyperbolic encounters.

{\it Simulations of eccentric orbits---}\label{sec:4/eccentric}%
We begin each evolution with a ``burn-in'' stage, where the motion of the scalar charge is fixed to a geodesic.   After $\sim\! 4$ orbits, the impact of the initial conditions (arbitrarily chosen to be zero) has decayed and we turn on 
the self-force term smoothly (during an apoapsis passage, where it is relatively small). Subsequently, the system is evolved self-consistently as described above. 

The left panel of Fig.~\ref{fig:ecc_xy} shows the orbit of a typical evolution, with $\epsilon = 0.02$ and initial apoapsis and periapsis at $r_+=100M$ and $r_-=5.26M$, respectively, corresponding to initial eccentricity $e:=(r_+-r_-)/(r_++r_-)=0.9$. %
The particle traces out $27$ orbits before it plunges into the central black hole. Both periapsis distance and periapsis advance change throughout the simulation, creating an irregular pattern. The worldtube radius of the depicted simulation was set according to $R_{6M} = 0.2M$. We additionally ran an evolution with $R_{6M} = 0.4M$, from which we estimate a total accumulated phase error of 0.24 radians up to the light ring crossing at $r_p = 3M$ (out of $\sim\! 170$ radians). In the supplemental material, we present convergence tests using different numerical resolutions of the DG scheme which show that we are accurately resolving the error induced by the worldtube.
The right panel displays a few multipole modes of the corresponding scalar-field waveform during the final stage of evolution, plotted against retarded KS time; we use notation whereby $\Psi_{lm}$ represents the real part of the  $(l,m)$ spherical-harmonic mode of $r \Psi / Q$, here extracted at $r=800M$.
The monopole $\Psi_{00}$ oscillates around a constant value with the radial  frequency.
Higher modes show more complex behavior due to strong precession and other relativistic effects.
The waveforms shown in Fig.~\ref{fig:ecc_xy} are extracted at a finite radius $r\!=\!800M$ and will therefore not exactly match the waveform at null infinity. An analysis of the incurred error is given in the supplemental material. Waveform extrapolation to null infinity is already available for the full gravitational case~\cite{Boyle:2019kee,Moxon:2021gbv}, so the absence of similar techniques for the results presented here does not affect the ultimate goal of our research.

\begin{figure}
	\includegraphics[width=\linewidth,trim=20 15 10 10,clip=true]{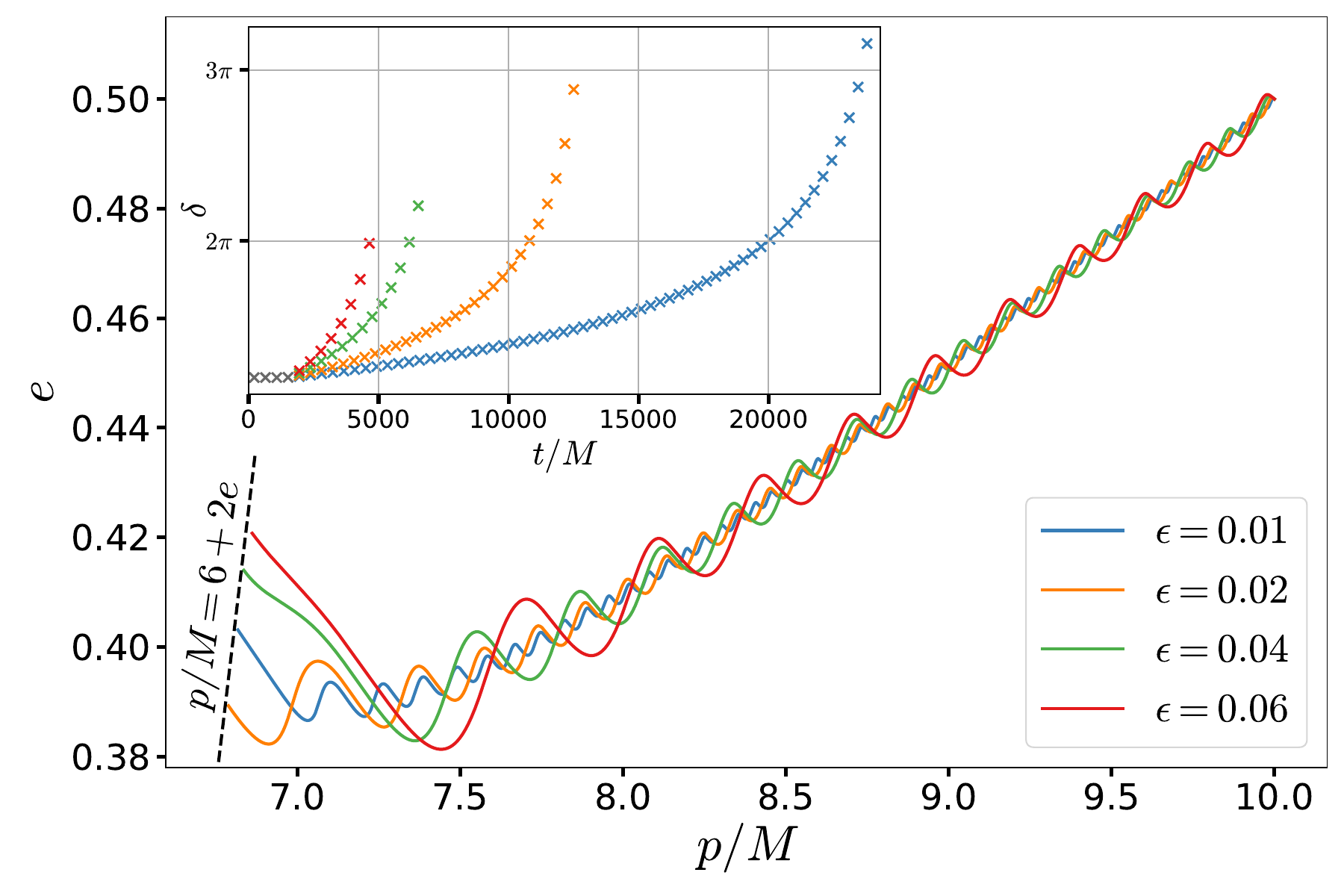}

	\caption{
		Eccentric inspirals for different strength-parameter $\epsilon$ of the self-force.
		The main panel plots eccentricity $e$ vs. semi-latus rectum $p$, where each system starts in the top right corner and moves toward smaller $p$.
	All simulations oscillate about approximately the same line in the $pe$ plane, with systems at larger $\epsilon$ exhibiting fewer oscillations of a larger amplitude during their faster inspiral.
		The $pe$ tracks end at the separatrix (dashed black) where geodesic orbits transition to trajectories plunging into the black hole. The inset shows the periastron advance $\delta$ of the simulations, measured in radians per radial cycle. The initial four grey points represent the burn-in stage where the particle's trajectory is fixed to a geodesic.
	}
	\label{fig:p_vs_e}
\end{figure}
The simulation used 2.3 million grid points distributed over 560 computational cores. Between simulation start and event-horizon crossing, it ran for $41000M$ with a wall time of 27 hours. While the scalar wave equation solved here is less computationally demanding than the full Einstein equations, these speeds provide a good indication of the power inherent in the worldtube excision method combined with a modern, scalable NR code.
Figure~\ref{fig:p_vs_e} shows the evolution of the two principal orbital elements---eccentricity $e$ and semi-latus rectum $p:=2r_+r_-/(r_+ + r_-)$---in a series of simulations with different values of the inspiral parameter $\epsilon$. All simulations start on the same initial geodesic with
$e\!=\!0.5$ and $p\!=\!10 M$ (corresponding to     apoapsis $20M$ and periapsis $6.66 M$)
  and end at the innermost stable orbit, $p/M\! =\! 6+ 2e$ (beyond which the $p,e$ parametrization ceases to be valid).
The radiative evolution decreases $p$ monotonically, and generally circularizes the orbit---except very near the plunge, where $e$ picks up again (a phenomenon familiar from previous calculations in black-hole perturbation theory \cite{Warburton:2011fk,VanDeMeent:2018cgn}).
The inset shows the evolution of periapsis advance $\delta$ (radians per radial period) for the same inspiral orbits.  
Once the self-consistent evolution begins (after the burn-in), the rate of periapsis advance increases monotonically. 
Just before the plunge, more than $5 \pi$ radians of azimuthal phase are traversed each radial period.

{\it Simulations of hyperbolic encounters---}\label{sec:4/hyperbolic}%
With modest adaptation, we can also apply our worldtube technique and code to unbound binaries, where the incident particle starts at infinite separation with velocity $v_\infty \geq 0$, and then either scatters back to infinity or gets captured by the black hole.
For unbound trajectories, we modify our burn-in stage. The particle's large initial separation from the BH means that we can initialize the scalar field with the leading-order expression for $\Psi^{\cal P}$ (the ``$1/$distance'', Coulomb-like field of an isolated scalar charge).
  We then evolve the scalar field by Eq.~(\ref{KG}) while forcing the particle to follow an incoming geodesic from its initial separation $r_p\!=\!200M$.  Around $r_p\!=\!100M$, we smoothly turn on the self-force terms, and subsequently proceed with the self-consistent evolution.
In the simulations presented here, the outer boundary of the numerical domain is placed at $r=1200M$.

\begin{figure*}
 \includegraphics[width=7.5cm,trim=25 20 0 10,clip=true]{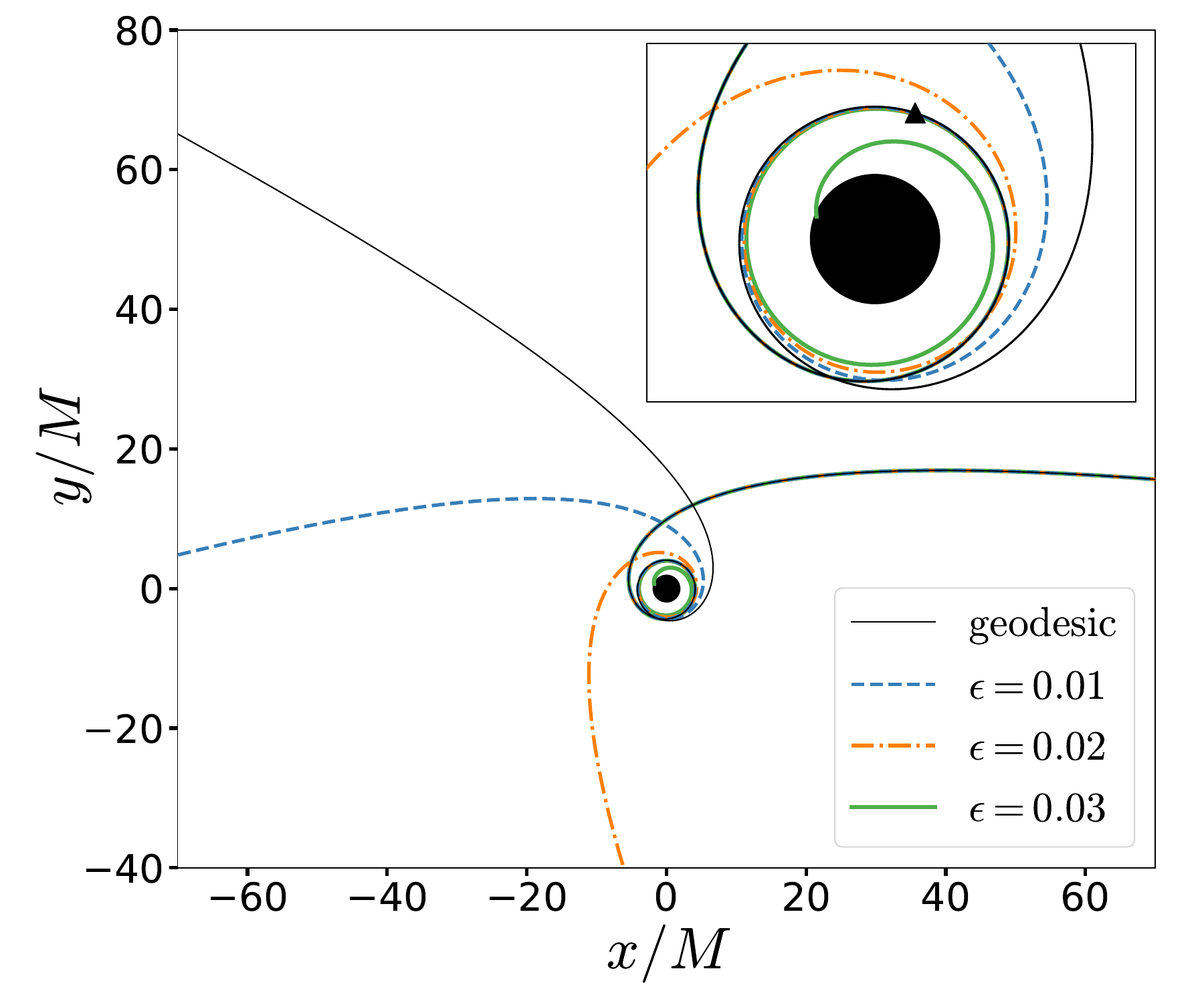}
  \includegraphics[width=10.3cm,trim=0 20 0 0,clip=true]{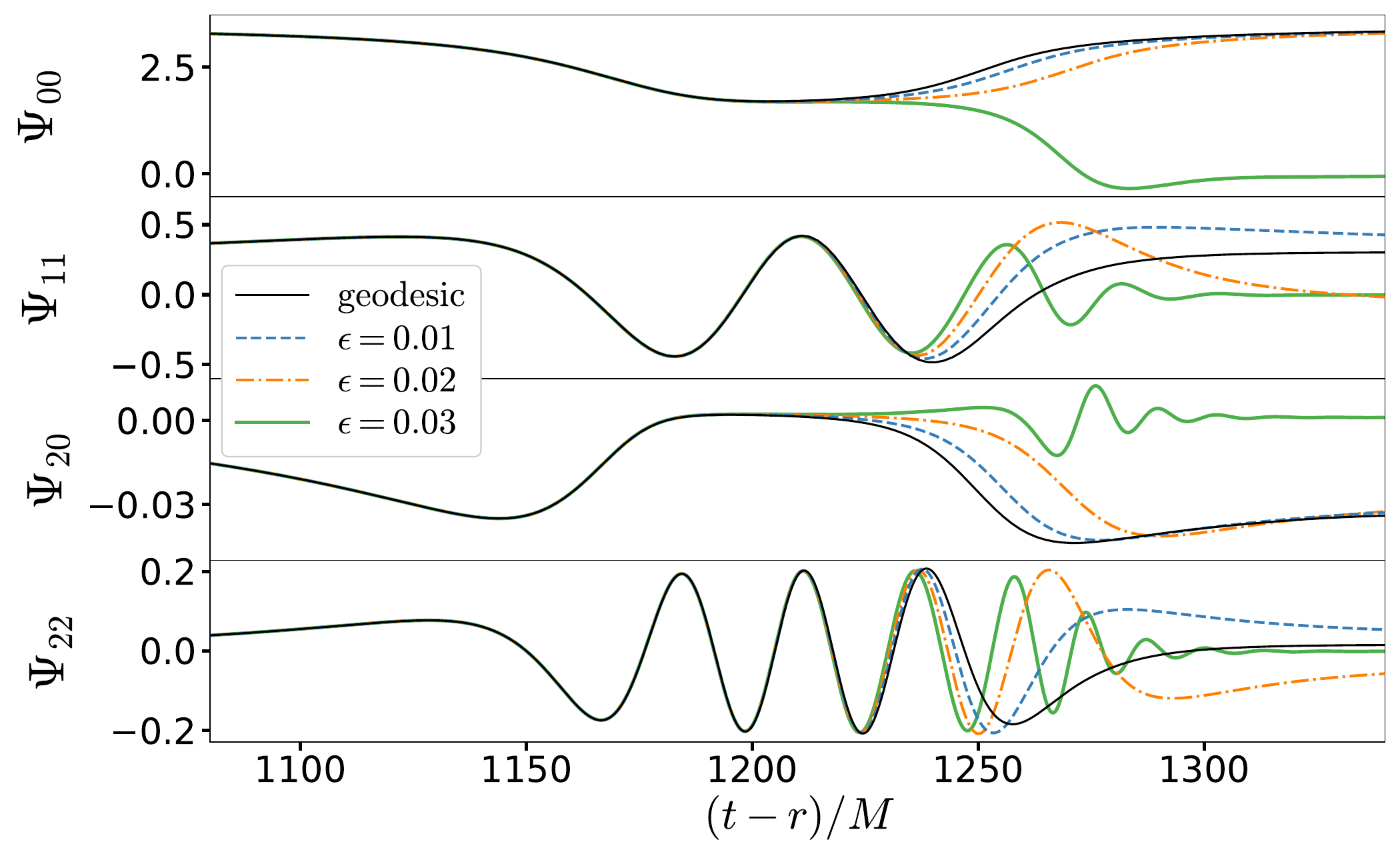}
\caption{Scattering/capture of a scalar charge around a black
    hole. \textbf{Left:} trajectories approaching from the
    positive x-axis, and separating near periapsis, marked by the black triangle.
    For $\epsilon = 0.01$ and $\epsilon = 0.02$, the self-force significantly increases the scattering angle compared to the geodesic.  For $\epsilon=0.03$, radiation removes enough energy for the
      particle to become bound and merge with the black hole.
      \textbf{Right:} plot of important waveform modes illustrating
      the qualitatively different behavior between scatter and capture.
      All simulations start on a geodesic with $v_\infty = 0.1$ and
      impact parameter $b = 40.21M$.
}
	\label{fig:hyper}
\end{figure*}

Figure~\ref{fig:hyper} displays a few representative simulations, all with $v_\infty\!=\! 0.1$ and impact parameter $b \!=\! 40.21 M$, corresponding to a scattering geodesic with parameters near the region where geodesics are captured by the BH.
The simulations differ only in the value of $\epsilon$, i.e.\ the strength of self-acceleration experienced by the particle.
For reference, we show also the geodesic case, with no self-force ($\epsilon=0$), which completes almost two full revolutions before scattering off to infinity (``zoom-whirl'' behavior). For $\epsilon>0$, the particle loses energy and angular momentum through scalar-wave emission and the scattering angle increases. 
The scattering angles in the simulations shown are approximately $12.29$ radians in the geodesic case, $13.08$ radians for $\epsilon = 0.01$, and $14.85$ radians for $\epsilon = 0.02$. Particularly interesting is the simulation with $\epsilon = 0.03$, in which  %
the charge loses enough energy to be captured by the BH. 
The inset enlarges the region around the central hole, showing how the four worldlines are almost identical as they approach (counter-clockwise) but begin to separate as they lose energy at different rates.

The worldtube radius of the simulations shown here is set as $R_{6M}\! =\! 0.4M$. We repeated each simulation with a larger worldtube radius of $R_{6M}\! =\! 0.8M$, allowing us to make error estimates. We find this changes the scattering angle by $0.1\%$ and $0.6\%$ for $\epsilon=0.01$ and $\epsilon=0.02$, respectively, indicating increased sensitivity near the threshold to capture. The error in the dissipated energy is more stable at 2.3\% and 2.9\%, respectively.

The right panel of Fig.~\ref{fig:hyper} shows the corresponding waveform modes, as extracted at $r=950M$.  
For $\epsilon=0.03$, the scattering signal is replaced with a rapid ringdown. These waveforms provide a stark illustration of sensitivity  near the capture threshold and the qualitative different outcomes on either side.

{\it Summary and outlook---}\label{sec:4/conclusions}%
We have presented a highly versatile implementation of the worldtube excision method for compact binary systems with disparate scales, showcasing its computational efficiency and flexibility with very long simulations in computationally challenging setups including highly-eccentric binaries and scattering orbits. Although we have confined ourselves to a model problem, there are encouraging conclusions to be drawn about the efficacy of the method in resolving scale disparity in actual BH binary simulations: replacing the scalar charge with a small black hole (at fixed $\epsilon=q$) would incur additional computational overhead from having to solve the full Einstein equations, but hardly more beyond that.

Our code also represents the first means of simulating generic, fully self-consistent self-forced binary evolutions. This can provide powerful benchmarking for perturbative approaches based on expansions in $\epsilon$ or on separations of time scales, such as the multiscale method underpinning current self-force waveform generation frameworks  for bound inspirals~\cite{LISAConsortiumWaveformWorkingGroup:2023arg} (as applied to scalar-field evolutions in Ref.~\cite{Speri:2024qak}), the self-force method of Ref.\ \cite{Barack:2022pde} for scattering orbits, or the multiple-scale transition-to-plunge analysis methods of Refs.~\cite{Kuchler:2024esj,Becker:2024xdi}.  We intend to use our code to pursue such analyses in forthcoming work.

We have also begun work to implement worldtube excision in binary BH simulations, our ultimate goal. In this pure-gravity case, we solve the full Einstein's equations in vacuum, with an excised worldtube inside which the metric is prescribed analytically using a model of a tidally perturbed black hole~\cite{Poisson:2005pi,Poisson:2009qj,Poisson:2018qqd,Yunes:2005ve,Chatziioannou:2016kem,Pani:2015hfa,LeTiec:2020bos}. The key step is the matching of the metric on the surface of the worldtube at each timestep.  Two approaches are being explored. One involves matching both the gauge and the a-priori unknown tidal deformation parameters of the internal BH.  Another, more elegant approach is based on matching a suitable set of curvature invariants. These methods will be fleshed out and numerically implemented in forthcoming work.

{\it Acknowledgements---}%
AP acknowledges the support of a Royal Society University Research Fellowship and the ERC Consolidator/UKRI Frontier Research Grant GWModels (selected by the ERC and funded by UKRI [grant number EP/Y008251/1]). This material is based upon work supported by the National Science Foundation under Grants No. PHY-2407742, No. PHY- 2207342, and No. OAC-2209655 at Cornell. Any opinions, findings, and conclusions or recommendations expressed in this material are those of the author(s) and do not necessarily reflect the views of the National Science Foundation. This work was supported by the Sherman Fairchild Foundation at Cornell. This work was supported in part by the Sherman Fairchild Foundation and by NSF Grants No.~PHY-2309211, No.~PHY-2309231, and No.~OAC-2209656 at Caltech. This work was supported in part by NSF awards PHY-2208014 and AST-2219109, the Dan Black Family Trust, and Nicholas and Lee Begovich at Cal State Fullerton. Computations were performed on the Urania HPC system at the Max Planck Computing and Data Facility.

SpECTRE uses \texttt{Charm++}/\texttt{Converse}~\cite{laxmikant_kale_2021_5597907,kale1996charm++}, which was developed by the Parallel Programming Laboratory in the Department of Computer Science at the University of Illinois at Urbana-Champaign. SpECTRE uses \texttt{Blaze}~\cite{Blaze1,Blaze2}, \texttt{HDF5}~\cite{hdf5}, the GNU Scientific Library (\texttt{GSL})~\cite{gsl}, \texttt{yaml-cpp}~\cite{yamlcpp}, \texttt{pybind11}~\cite{pybind11}, \texttt{libsharp}~\cite{libsharp}, and \texttt{LIBXSMM}~\cite{libxsmm}. The figures were produced with \texttt{matplotlib}~\cite{Hunter:2007, thomas_a_caswell_2020_3948793}, \texttt{NumPy}~\cite{harris2020array}, and \texttt{ParaView}~\cite{paraview,paraview2}.

\bibliography{references}

\end{document}